\documentclass[%
 reprint,
superscriptaddress,
 amsmath,amssymb,floatfix,
]{revtex4-2}

\usepackage{graphicx}
\usepackage{dcolumn}
\usepackage{comment}
\usepackage{bm}
\usepackage{color}

\begin{document}
\title{Interplay between an absorbing phase
transition and synchronization in a driven granular system}
\author{R. Maire}
\affiliation{Universit\'e Paris-Saclay, CNRS, Laboratoire de Physique des Solides, 91405 Orsay, France}
\author{A. Plati}
\affiliation{Universit\'e Paris-Saclay, CNRS, Laboratoire de Physique des Solides, 91405 Orsay, France}
\author{M. Stockinger}
\affiliation{Universit\'e Paris-Saclay, CNRS, Laboratoire de Physique des Solides, 91405 Orsay, France}
\affiliation{Max Planck Institute for Gravitational Physics (Albert Einstein Institute), Am Mühlenberg 1, 14476 Potsdam, Germany}
\author{E. Trizac}
\affiliation{LPTMS, UMR 8626, CNRS, Universit\'e Paris-Saclay, 91405
  Orsay, France}
\affiliation{Ecole normale sup\'erieure de Lyon, F-69364 Lyon, France}
\author{ F. Smallenburg}
\affiliation{Universit\'e Paris-Saclay, CNRS, Laboratoire de Physique des Solides, 91405 Orsay, France}
\author{G. Foffi}
\affiliation{Universit\'e Paris-Saclay, CNRS, Laboratoire de Physique des Solides, 91405 Orsay, France}
\date{\today}

\begin{abstract}
Absorbing phase transitions (APTs) are widespread in non-equilibrium systems, spanning condensed matter, epidemics, earthquakes, ecology, and chemical reactions. APTs feature an absorbing state in which the system becomes entrapped, along with a transition, either continuous or discontinuous, to an active state.  Understanding which physical mechanisms determine the order of these transitions
represents a challenging open problem in non-equilibrium statistical mechanics.  Here, by numerical simulations and mean-field analysis, we show that a quasi-2d vibrofluidized granular system exhibits a novel form of APT. 
The absorbing phase is observed in the horizontal dynamics below a critical packing fraction, and can be continuous or discontinuous based on the emergent degree of synchronization in the vertical motion. Our results provide a direct representation of a feasible experimental scenario, showcasing a surprising interplay between dynamic phase transition and synchronization.
\end{abstract}

\maketitle


In non-equilibrium statistical physics, an \textit{absorbing phase transition} (APT) happens when a system transitions from a steady diffusive state to a static or periodic state with no further evolution upon varying a control parameter. 
The characteristics and universality class of these transitions have been extensively discussed in the literature~\cite{hinrichsen2000non, lubeck2004universal}. APTs have been observed in many different contexts, from epidemiology~\cite{Mata2021} to sandpile models~\cite{Dickman1998}, from quantum systems~\cite{marcuzzi2016absorbing} to chemical reactions~\cite{ziff1986kinetic} and,  since the  pioneering work by Pine and coworkers~\cite{pine2005chaos}, particle systems have become a major area of study for APTs. Specifically, APTs have been observed in low Reynolds number reversible  suspensions~\cite{pine2005chaos, corte2008random, nagamanasa2014experimental, jeanneret2014geometrically},  dense colloidal systems~\cite{keim2014mechanical,tjhung2015hyperuniform}, emulsions~\cite{knowltonsm14, Weijs2015}, liquid crystals~\cite{takeuchi2007directed}, glasses~\cite{fiocco2013,regev2013onset, bhaumik2021role}, jammed systems~\cite{tjhung2015hyperuniform, Ness2020}, granular materials~\cite{Mobius2014,neel2014dynamics}  and turbulent systems~\cite{chantry2017universal}.  In several of the cases mentioned above~\cite{pine2005chaos, corte2008random, nagamanasa2014experimental, takeuchi2007directed, Ness2020, chantry2017universal}, the transition occurs in a continuous manner, and this behavior has been interpreted within the framework of directed percolation (DP)~\cite{Rossi2000} or conserved directed percolation (CDP) models, such as the Manna models~\cite{LeDoussal2015, Martiniani2019}. In many other situations~\cite{Kawasaki2016, jeanneret2014geometrically, leishangthem2017yielding}, however, it has been found that the transition is discontinuous, and it has been argued that interactions, possibly arising from hydrodynamics or elasticity, might be responsible for this effect~\cite{mari2022absorbing}.

In this letter we investigate a simple vibrofluidized granular model in which macroscopic spherical beads, confined within a quasi-2D cell, undergo an APT from a state in which they are locked in a vertical motion to a diffusive and active state in the horizontal direction. Interestingly, by adjusting parameters like the confinement height or the vibration amplitude, we can induce either a continuous or discontinuous transition.
While previous studies have identified systems exhibiting both types of transitions~\cite{Fisher97, van1998wilson,assis2009discontinuous,Jagla2014, mari2022absorbing,lorenzana2023emergent}, the underlying physical explanations have often remained unclear. Here, we demonstrate that a synchronization mechanism between grains in the vertical direction provides a clear understanding of this dual behavior within our system. Our observations are obtained by simulations on both a realistic and a coarse-grained model solved by molecular dynamics. The results are rationalized using kinetic theory.

\begin{figure}
\centering
\includegraphics[width=0.99\columnwidth,clip=true]{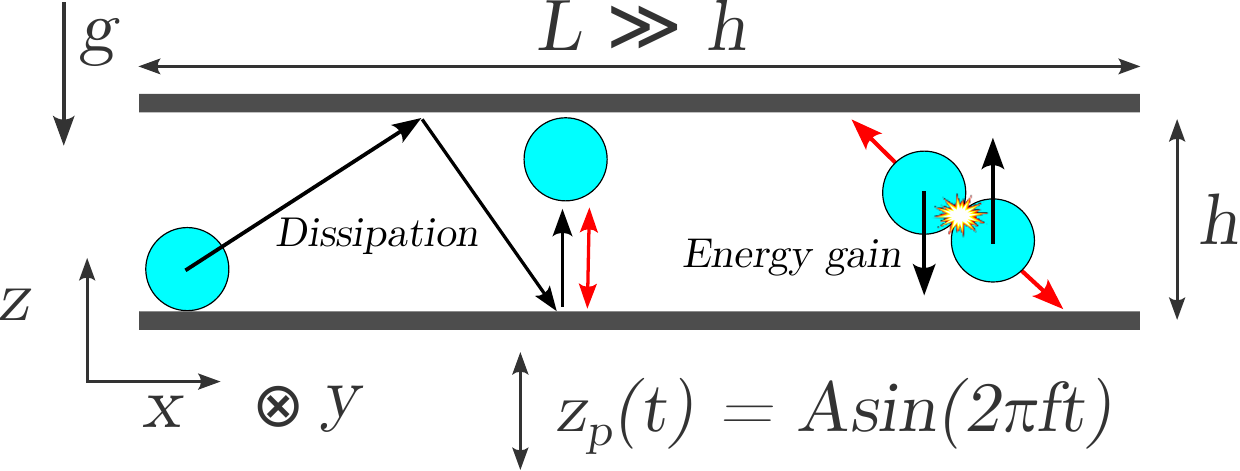}

\caption{Numerical quasi-2d geometry used in realistic simulations. A vertical displacement $z_p(t)$ is imposed on the box in order to provide external energy to the system. Because of tangential frictional forces, the grains lose horizontal energy during collisions with the top and bottom walls. During non-planar collisions between grains, the vertical energy gained from the vibration of the plate is transferred to the $xy$ components of the velocities of the particles.}
\label{fig:Fig1}
\end{figure}

\emph{Realistic model--} Our first model consists of  a vertically vibrating  granular system confined between two plates in a quasi-2D geometry (see Fig. \ref{fig:Fig1}). The evolution is studied using molecular dynamics simulations based on the Discrete Element Method (DEM) \cite{Cundall79,PoeschelBook} using LAMMPS \cite{Plimpton2022, kloss2010granular}. This numerical approach is widely adopted to model realistic setups in-silico \cite{Plati2019,Osullivan2009,Plati2024,Plati2022}, utilizing precise contact mechanics models \cite{Zhang2005,PoeschelBook,PopovBook} for the grain-grain and grain-plate interactions (see SI).   
The simulation box has a square base of side $L$ in the $xy$-plane and height $h$ with $L \gg h$. Particles in the box are confined in the $z$-direction by two horizontal plates. Note that the grain-plate interaction also includes tangential friction, which is crucial for the phenomenon under study. In the $x$ and $y$ directions we impose periodic boundary conditions.
The box vibrates in the $z$-direction following a sinusoidal motion with frequency $f$ and amplitude $A$ and is filled with monodisperse spherical grains of spatial coordinates $\{x,y,z\}$, translational velocities $\{v_x,v_y,v_z\}$, rotational velocities $\{\omega_x,\omega_y,\omega_z\}$  diameter $\sigma$ and mass $m$. 
The physical parameters, which are fixed for all the DEM simulations presented in this paper, are: $\sigma=2.5$ mm, $f=53$ Hz, $m=6.54\times 10^{-5}$ kg. The number $N$ of simulated grains ranges from $10^{3}$ to $2\times10^{4}$ depending on the specific analysis.
In the following, we will consider lengths expressed in terms of grain diameter $\sigma$, times in terms of inverse driving frequency $1/f$, masses in terms of grain mass $m$. Thus, the energy unit is given by  $m(\sigma f)^2=1.15\times 10^{-6}$ J.

In this system, collisions with the upper and lower plates inject kinetic energy along the $z$ direction, but dissipate any kinetic energy in the $xy$ plane due to tangential frictional forces. Hence, an isolated particle with a finite velocity in the $xy$ plane will slow down and eventually come to a stop with respect to its horizontal motion, while continuing to vertically bounce between the two vibrating plates. However, collisions between grains allow for $z$-to-$xy$ energy conversion (see Fig.~\ref{fig:Fig1}). 
The resulting  dynamics of an isolated grain follow two possible scenarios: it can either experience a collision-free flight and lose all its horizontal energy, or collide again gaining new horizontal energy.

\begin{figure}
\includegraphics[width=0.99\columnwidth,clip=true]{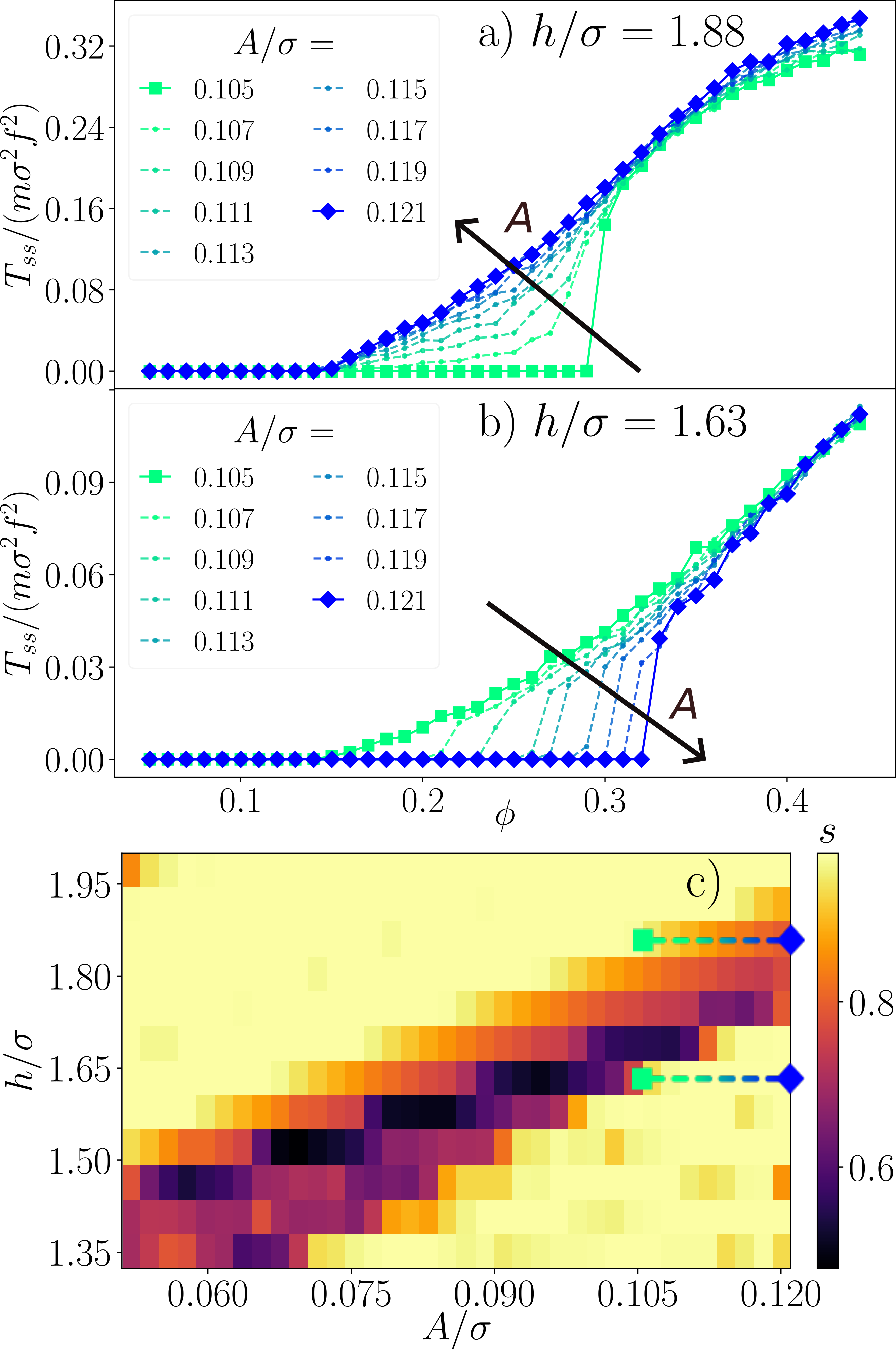}
\centering
\caption{Horizontal kinetic energy in the steady state as a function of $\phi$ for $h=1.88\sigma$ (a), $h=1.63\sigma$ (b) and different vibration amplitudes. c) Synchronization map in the $\{A,h\}$ parameter space. We note that the degree of synchronization $s$ exhibits a non-monotonous behaviour as a function of both $A$ and $h$. Simulations are performed with $N=10^{3}$ grains. }
\label{fig:Fig2}
\end{figure}

\emph{APT and synchronization--} DEM simulations are initialized by placing $N$ grains with random velocities in the vibrating quasi-2D box. After a transient, the dynamics reach a non-equilibrium steady state  where particles bounce between the confining plates. In this situation,  particles are either mobile or immobile in the $xy$-plane, depending on the system parameters (see Video1). 
We characterize the granular horizontal dynamics 
with the mean horizontal kinetic energy of the grains: 
\begin{equation}
T=\frac{1}{\mathcal{T}}\int_0^{\mathcal{T}}dt\frac{m}{2}\langle v_x^2(t) +v_y^2(t) \rangle
\label{eq:Tgranular}    
\end{equation}
where $\langle \cdot \rangle$ refers to the average over all the particles in an instantaneous configuration and $\mathcal{T}$ is the observation time. This quantity is usually called granular temperature. 
In Figs. \ref{fig:Fig2}ab we show the steady state value $T_{ss}$ of the granular temperature as a function of the system's two-dimensional packing fraction $\phi=\pi N\sigma^2/4L^2$ for different combinations of amplitude $A$ and box height $h$. We make two main observations. First, all the curves show an APT between an absorbing state ($T_{ss}=0$) below a critical packing fraction $\phi_c$ and an active state ($T_{ss}>0$) above $\phi_c$ (see Video 2). Second, The APTs can be either continuous or discontinuous depending on the specific combination of $A$ and $h$.

The first observation can be explained as follows.
At low packing fractions ($\phi<\phi_c$), where collisions between grains are rare, the tangential friction during grain-plate collisions dissipates all the energy in the $xy$-directions, leading to dynamical arrest. 
At moderate packing fractions ($\phi>\phi_c$), where the grain-grain collision frequency is higher, energy transfer from the $z$-direction to the $xy$-plane keeps the system at a finite horizontal kinetic energy.
The second observation requires a more detailed investigation.
In Fig.~\ref{fig:Fig2}a, we show the behavior of $T_{ss}$ for different amplitudes with a fixed $h=1.88\sigma$. Here we see that, increasing $A$, the transition goes from discontinuous to continuous and the critical packing fraction decreases. The same analysis performed for $h=1.63\sigma$ (Fig.~\ref{fig:Fig2}b) shows an opposite trend: upon increasing $A$, we observe a change from a continuous to a discontinuous transition and an increase of the critical packing fraction. By comparing the two sets of curves highlighted with larger symbols in both panels, we also note that
raising $h$ at fixed amplitude implies going from a continuous to a discontinuous transition with an increase of $\phi_c$ for $A=0.105\sigma$ while for  $A=0.121\sigma$ the opposite happens.
We point out that, regardless of the variation of $A$ or $h$, passing from a second-order-like to a first-order-like transition is always accompanied by an increase of $\phi_c$.
From this phenomenology, it is clear that the effects of $A$ and $h$ on the APT cannot be rationalized separately. 

To explain this unexpected behavior, we examine the vertical motion in the absorbing state (low $\phi$) for different $A$ and $h$. 
Vertical trajectories reveal that, depending on the choice of $A$ and $h$, grains can be either in a synchronized  or in an asynchronized state (see also Video3 in the S.I.). During the synchronized motion, vertical trajectories of the grains are periodic, with all particles showing the same period and phase after a short initial synchronization time. Hence, at each instant $t$, they all have the same $z(t)$ and $v_z(t)$. Asynchronous motion, on the other hand, is observed in two different forms: grains can move aperiodically or periodically with the same period but different phases. We point out that synchronization phenomena have already been observed in vibrated granular systems \cite{10.1007/978-3-319-27965-7_32, schindler2023phase,rivas2011sudden,Pieraski1983,Pieraski1985,Chastaing2015} and can be understood theoretically by means of simple dynamical models \cite{everson1986chaotic, Metha1990,Luck1993}.
We define the degree of synchronization of the system as
    $s = \lim_{t_0,t\to \infty}\frac{1}{t}\int_{t_0}^{t_0+t}dt'|\langle e^{i\theta^j (t')}\rangle|$,
where 
$\theta^j$ is the effective phase of particle $j$ (see SI). 
Measuring $s$ as a function of $A$ and $h$ we obtain the synchronization map shown in Fig.~\ref{fig:Fig2}c.
Here we can see that the degree of synchronization is determined by the interplay between shaking amplitude and box height in a non-trivial way. In particular, upon varying either $A$ or $h$, it is possible to see the system transition from synchronized to asynchronized and back without changing any other parameters.
Based on the synchronization of isolated particles, we can look at the horizontal motion from a new point of view. To this aim, we highlight in Fig.~\ref{fig:Fig2}c the path in $\{A,h\}$ that is followed in Figs.~\ref{fig:Fig2}ab. We can conclude that, for the $xy$ motion, vertical synchronization leads to discontinuous transitions, while vertical asynchronization leads to continuous ones. A first qualitative insight on the role of synchronization comes from the fact that it can affect the energy gain in a grain-grain collision. Indeed, two synchronized particles will collide while being at the same height (in-plane collisions), and hence experience a less efficient $z$-to-$xy$ energy transfer with respect to what happens when grains collide not being at the same height (off-plane collisions). 
Hence, vertical synchronization of the particles can severely reduce the injection of horizontal kinetic energy into the system promoting the transition to the absorbing state. 

Summarizing, the observed phenomenology is attributable to the interplay between two distinct transitions: one between absorbing and active states in the $xy$-motion, and the other between synchronized and asynchronized states in the $z$-dynamics.

\emph{Coarse grained model--}
 In order to reduce the complexity of this phenomenology to its essential ingredients, we introduce a simplified  2D coarse-grained model that incorporates the three primary mechanisms derived from the analysis of the realistic model: energy injection at collision, slowdown of the velocity during the ``free flight" and synchronization. 
 
 The beads are modeled by identical hard disks in 2D of mass $m$ and diameter $\sigma$ undergoing active collisions \cite{brito2013hydrodynamic} characterized by a dissipative coefficient of restitution $ \alpha$ and a velocity injection $\Delta_{ij}$ between particles $i$ and $j$ which accounts for the $z$ to $xy$ energy transfer of the DEM model. This is expressed by the collision rule:
\begin{equation}
    \begin{split}
        \mathbf v_i'&= \mathbf v_i + \dfrac{1+\alpha}{2}(\mathbf v_{ij}\cdot \hat{\bm\sigma}_{ij})\hat{\bm\sigma}_{ij} + \Delta_{ij} \hat{\bm\sigma}_{ij} \\
        \mathbf v_j'&= \mathbf v_j - \dfrac{1+\alpha}{2}(\mathbf v_{ij}\cdot \hat{\bm\sigma}_{ij})\hat{\bm\sigma}_{ij} - \Delta_{ij} \hat{\bm\sigma}_{ij},
    \end{split}
    \label{eq: collRule}
\end{equation}
where $0\leq\alpha\leq1$ is the coefficient of restitution, $\Delta_{ij}>0$ is the velocity injection, $\mathbf v_i'$ the post-collision velocity of particle $i$, $\mathbf v_i$ its pre-collision velocity, and $\mathbf v_{ij}$ and $\hat{\bm\sigma}_{ij}$ are respectively the relative velocity between particles $i$ and $j$ and the unit vector joining them. The $\Delta$ imitates the energy transfer from vertical component to horizontal found in the realistic model.
A viscous drag during the free flight is introduced to reflect the dissipative collisions with the plates in the realistic model:
\begin{equation}
    \dfrac{\mathrm{d} \mathbf v_i}{\mathrm{d} t} = -\gamma \mathbf v_i.
\end{equation}

The final ingredient of the model is synchronization, which is introduced explicitly through a history-dependent $\Delta_{ij}$ and a synchronization time $\tau_s$:
\begin{equation}
    \Delta_{ij} = \left\{
    \begin{array}{ll}
        \Delta > 0 & \mbox{if  } \delta t_i< \tau_s \mbox{  or  } \delta t_j < \tau_s\\
        0 & \mbox{otherwise,}
    \end{array}
\right.
\label{eq: vdependentDelta}
\end{equation}
with $\delta t_i$ the time since the last collision of particle $i$. The idea is that synchronization arises on timescale $\tau_s$ (see S.I.) that is in competition with the typical collision time between particles. Synchronized particles will experience purely dissipative collisions.
In particular, the limit $\tau_s\rightarrow\infty$ corresponds to a system where no synchronization takes place.
A summary of the differences between the coarse-grained model and the realistic model is given in the S.I.

\begin{figure}[!ht]
\centering
\includegraphics[width=0.97\columnwidth,clip=true]{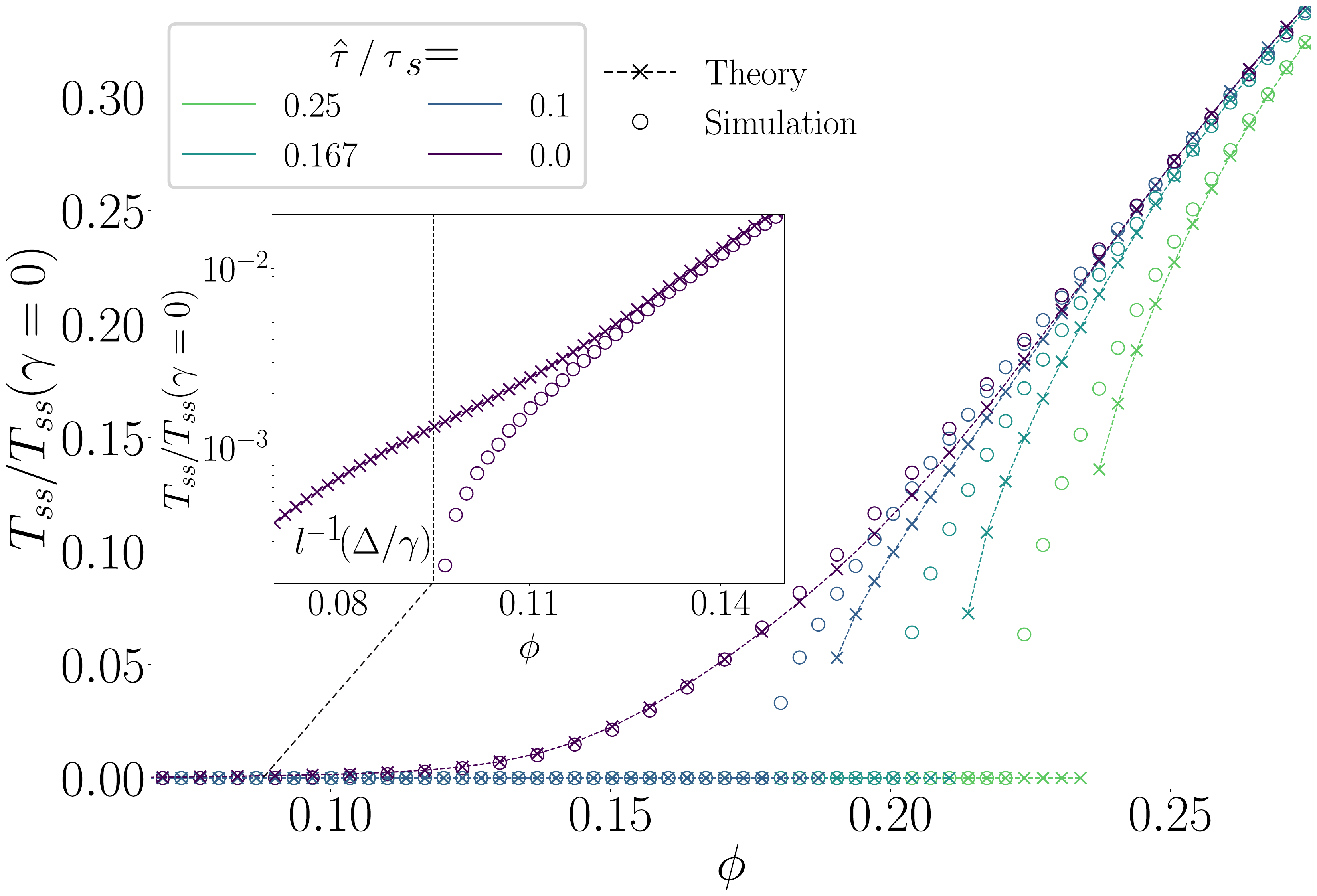}
\caption{Comparison between the theory and simulations. Effect of synchronization over the nature of transition and the critical packing fraction: $N = 20000$, $\hat \tau\Delta/\sigma = 0.025$, $\alpha = 0.95$ and $\hat \tau\gamma = 0.01$. The inset is a semi-log window of the small density behavior of the theory and simulation for the case without synchronization. The dashed vertical line represents the critical packing fraction predicted from the mean free path argument.} \label{fig:fig3}
\end{figure}

We perform event-driven simulations \cite{smallenburg2022efficient} of this model. Results are presented in Fig. \ref{fig:fig3}.  The behavior observed with the realistic model is recovered: when the synchronization time is finite a discontinuous transition is observed. In contrast, similar to the behavior found in Ref. \cite{lei2019hydrodynamics}, a continuous transition is found (see inset) when $\tau_s\rightarrow\infty$.
We also observed that, in the first case, the system reaches an absorbing state due to dissipative collisions while in the second case it is reached because of  the viscous drag.
We confirmed the nature of both transitions with a finite size analysis  (See S.I).

\emph{Theory--}
Following the approach presented in Ref.~\onlinecite{brito2013hydrodynamic}, the temperature change of the system, assuming homogeneity, is only determined by the energy injection and dissipation following the relation
\begin{equation}
    \dfrac{\partial T}{\partial t} = G(\phi, T) =\dfrac{\omega(\phi, T)}{2}\langle E' - E\rangle_{coll} - 2\gamma T,
    \label{eq: Tevolution}
\end{equation}
where $G(\phi, T)$ represents the rate of energy change due to collisions and drag, $\omega(\phi, T)$ is the frequency of collision and $\langle \dots \rangle_{coll}$ is an average over collisions \cite{pagonabarraga2001randomly} (see S.I.).

An exact calculation of the collisional average in Eq.~\ref{eq: Tevolution}  is challenging due to the cumbersome form of $\Delta_{ij}$ (Eq.~\ref{eq: vdependentDelta}), however we can make a reasonable estimate  by assuming that the history dependent $\Delta_{ij}$ can be replaced by an effective one: $\overline \Delta$. We propose:  
\begin{equation}
    \overline{\Delta}(\phi, T, \tau_s)\equiv\overline{\Delta}_{ij} = \Delta \left(1 - e^{-2\omega(\phi, T)\tau_s}\right),
    \label{eq: approx one}
\end{equation}
where the term in parenthesis 
is the probability that at least one of the two particles involved in the collision has collided at a time smaller than $\tau_s$ in the past, assuming uncorrelated particles and Poissonian collisions \cite{visco2008non}.

Finally,  we simplify the averages over collisions including a $\Delta_{ij}$ by doing the following approximation when calculating the average 
$\langle \Delta_{ij}g(\mathbf v_i, \mathbf v_j)\rangle_{coll} \simeq \overline{\Delta}\langle g(\mathbf v_i, \mathbf v_j)\rangle_{coll}$ with  $g(\mathbf v_i, \mathbf v_j)$ an arbitrary function. This approximation essentially neglects any correlations between particle velocities and their synchronization state. Note that this approximation becomes exact in the limit $\tau_s\rightarrow\infty$ since $\Delta_{ij} = \Delta$, a constant, in this case.

Assuming molecular chaos and a Gaussian velocity distribution, Eq. \ref{eq: Tevolution} reduces to (see S.I.):
\begin{equation}
    G(\phi, T) = \frac{\omega(\phi, T)}{2}(m\bar\Delta^2+\alpha\bar\Delta\sqrt{\pi m T}-T(1-\alpha^2))-2\gamma T.
    \label{eq: GsimpleDelta}
\end{equation}
In the limit $\tau_s\rightarrow\infty$ and by assuming Enskog's frequency of collision for $\omega(\phi,T)$ (see S.I.), a non trivial zero of Eq. \ref{eq: GsimpleDelta} can be found exactly and leads to the following steady state temperature:
\begin{equation}
    T_{ss} = \left(\dfrac{\epsilon  + \sqrt{\epsilon^2 +4m\Delta^2(1-\alpha^2)}}{2(1-\alpha^2)}\right) ^2
    \label{eq: steadtstatetemp}
\end{equation}
with $\epsilon= \alpha\Delta \sqrt{\pi m}-4\gamma/\tilde\omega$ and $\tilde \omega = \frac{ 8 \phi \chi(\phi)}{\sigma\sqrt{\pi m}}$ where $\chi$ is the Enskog factor taken to be the radial pair distribution function at contact of an equilibrium system at the same density. 
Eq. \ref{eq: steadtstatetemp} always predicts a stable finite temperature at finite $\phi$. That is, it does not predict the existence of a continuous transition as observed in the simulations. This is caused by our frequency of collision taken to be the same as an equilibrium one. This assumption only holds as long as the mean free time is small compared to $1/\gamma$. Indeed, we can show that below a packing fraction $\phi_c$, the system must come to a rest. At this $\phi_c$, the temperature should vanish and hence the exiting velocity of a particle after a collision must be equal to $\Delta$ which allows it to travel a distance $\Delta/\gamma$. By equating this distance with the mean free path \cite{lei2019hydrodynamics} $l(\phi)$ we obtain an equation for $\phi_c$: $l(\phi_c) = \Delta/\gamma$.  In the inset of Fig. \ref{fig:fig3}, we plot a vertical line corresponding to this value, which agrees well with the transition packing fraction observed in our simulations in the absence of synchronization.
In Fig. \ref{fig:fig3}, we show a comparison between simulations of the coarse-grained model and the theory. Without synchronization ($1/\tau_s = 0$), the theory given by Eq. \ref{eq: steadtstatetemp} works very well, except at the transition as seen in the inset due to the breakdown of the assumption on the frequency of collision. When the synchronization time is finite, a numerical solution of the theory predicts well the transition point and is in good agreement with the numerical values from the simulation. Better agreement is hindered by the approximation done in Eq. \ref{eq: approx one} and by the hypothesis of homogeneity.
Overall, our kinetic theory clarifies that the physical mechanism underlying the occurrence of discontinuous APTs is the weakening of energy transfer at collisions caused by synchronization.

\emph{Conclusion--} Summarizing, we have explored both numerically  and theoretically  a vibrated  granular system confined in a quasi 2D geometry  that exhibits both a continuous and a discontinuous absorbing phase transition and we have demonstrated that the character of this transition is intricately linked to a synchronization mechanism occurring due to the confinement and hence the dominant dissipation mechanism.
Our work presents an interesting effect of synchronization on APTs that allows to reconcile the two different types of phase transition observed in non-equilibrium phenomena. Indeed, by adjusting the driving parameters, we have demonstrated the ability to shift the  order of the transition from continuous to discontinuous. 
Our model draws inspiration from a setup for vibrated granular matter which has been extensively investigated
\cite{rivas2011sudden, schindler2023phase, khain2011hydrodynamics, brito2013hydrodynamic, reyes2008effect, PhysRevLett.81.4369, PhysRevE.59.5855, PhysRevE.70.050301, PhysRevLett.89.044301, clerc2008liquid}, and can be readily implemented in future experimental studies of this phenomenon. 

\begin{acknowledgments}

We thank Emanuela Del Gado, Andrea Puglisi, Alberto Rosso  and Srikanth Sastry for carefully reading and commenting the manuscript.
This work has been done with the support of Investissements d'Avenir of LabEx PALM (Grant No. ANR-10-LABX-0039-PALM). 

\end{acknowledgments}

\bibliographystyle{apsrev4-1} 
\bibliography{granular}

\end{document}